\documentclass[8.5pt,twoside,twocolumn]{article}
\usepackage[super,sort&compress,comma]{natbib} 
\usepackage{mhchem}
\usepackage{times,mathptmx}
\usepackage{sectsty}
\usepackage{balance} 
\usepackage{amsmath} 
\usepackage{amssymb} 
\usepackage{graphicx} 
\usepackage{lastpage}
\usepackage[format=plain,justification=raggedright,singlelinecheck=false,font=small,labelfont=bf,labelsep=space]{caption} 
\usepackage{fancyhdr}
\usepackage{xcolor}
\begin{document}

\thispagestyle{plain}
\fancypagestyle{plain}{
\renewcommand{\headrulewidth}{1pt}}
\renewcommand{\thefootnote}{\fnsymbol{footnote}}
\renewcommand\footnoterule{\vspace*{1pt}%
\hrule width 3.4in height 0.4pt \vspace*{5pt}} 
\setcounter{secnumdepth}{5}

\makeatletter 
\def\subsubsection{\@startsection{subsubsection}{3}{10pt}{-1.25ex plus -1ex minus -.1ex}{0ex plus 0ex}{\normalsize\bf}} 
\def\paragraph{\@startsection{paragraph}{4}{10pt}{-1.25ex plus -1ex minus -.1ex}{0ex plus 0ex}{\normalsize\textit}} 
\renewcommand\@biblabel[1]{#1}            
\renewcommand\@makefntext[1]%
{\noindent\makebox[0pt][r]{\@thefnmark\,}#1}
\makeatother 
\renewcommand{\figurename}{\small{Fig.}~}
\sectionfont{\large}
\subsectionfont{\normalsize} 

\fancyfoot{}
\fancyfoot[RO]{\footnotesize{\sffamily{1--\pageref{LastPage} ~\textbar  \hspace{2pt}\thepage}}}
\fancyfoot[LE]{\footnotesize{\sffamily{\thepage~\textbar\hspace{3.45cm} 1--\pageref{LastPage}}}}
\fancyhead{}
\renewcommand{\headrulewidth}{1pt} 
\renewcommand{\footrulewidth}{1pt}
\setlength{\arrayrulewidth}{1pt}
\setlength{\columnsep}{6.5mm}
\setlength\bibsep{1pt}

\twocolumn[
  \begin{@twocolumnfalse}
\noindent\LARGE{\textbf{Particle diffusion in active fluids is non-monotonic in size}}

\vspace{0.6cm}

\noindent\large{\textbf{Alison E. Patteson,\textit{$^{a}$} Arvind Gopinath,\textit{$^{a,b}$} Prashant K. Purohit,\textit{$^{a}$} and
Paulo E. Arratia\textit{$^{a}$}}}\vspace{0.5cm}

\noindent\textit{\small{\textbf{Received Xth XXXXXXXXXX 20XX, Accepted Xth XXXXXXXXX 20XX\newline
First published on the web Xth XXXXXXXXXX 200X}}}

\noindent \textbf{\small{DOI: 10.1039/b000000x}}
\vspace{0.6cm}

\noindent \normalsize{We experimentally investigate the effect of particle size on the motion of
passive polystyrene spheres in suspensions of {\it Escherichia coli}. Using particles covering a range of sizes from 0.6
to 39 microns, we probe particle dynamics at both short and long time scales.
In all cases, the particles exhibit super-diffusive ballistic behavior at short
times before eventually transitioning to diffusive behavior. Surprisingly, we find a regime in which larger particles can diffuse faster than smaller particles: the particle
long-time effective diffusivity exhibits a peak in particle
size, which is a deviation from classical
thermal diffusion. We also find that the active contribution to particle diffusion is controlled by a dimensionless parameter, the P\'{e}clet number. A minimal model qualitatively explains the existence of the effective diffusivity peak and its dependence on bacterial
concentration. Our results have broad implications on characterizing active
fluids using concepts drawn from classical thermodynamics.}
\vspace{0.5cm}
 \end{@twocolumnfalse}
  ]

\section{Introduction}
\footnotetext{\dag~Electronic Supplementary Information (ESI) available: [details of any supplementary information available should be included here]. See DOI: 10.1039/b000000x/}


\footnotetext{\textit{$^{a}$~Department of Mechanical Engineering \& Applied Mechanics,
University of Pennsylvania, Philadelphia, PA 19104. Fax: XX XXXX XXXX; Tel: XX XXXX XXXX; E-mail: xxxx@aaa.bbb.ccc}}
\footnotetext{\textit{$^{b}$~School of Engineering, University of California Merced, Merced,
CA 95343.}}


\footnotetext{\ddag~Additional footnotes to the title and authors can be included \emph{e.g.}\ `Present address:' or `These authors contributed equally to this work' as above using the symbols: \ddag, \textsection, and \P. Please place the appropriate symbol next to the author's name and include a \texttt{\textbackslash footnotetext} entry in the the correct place in the list.}

\begin{figure}[]
\centering
  \includegraphics[width=\columnwidth]{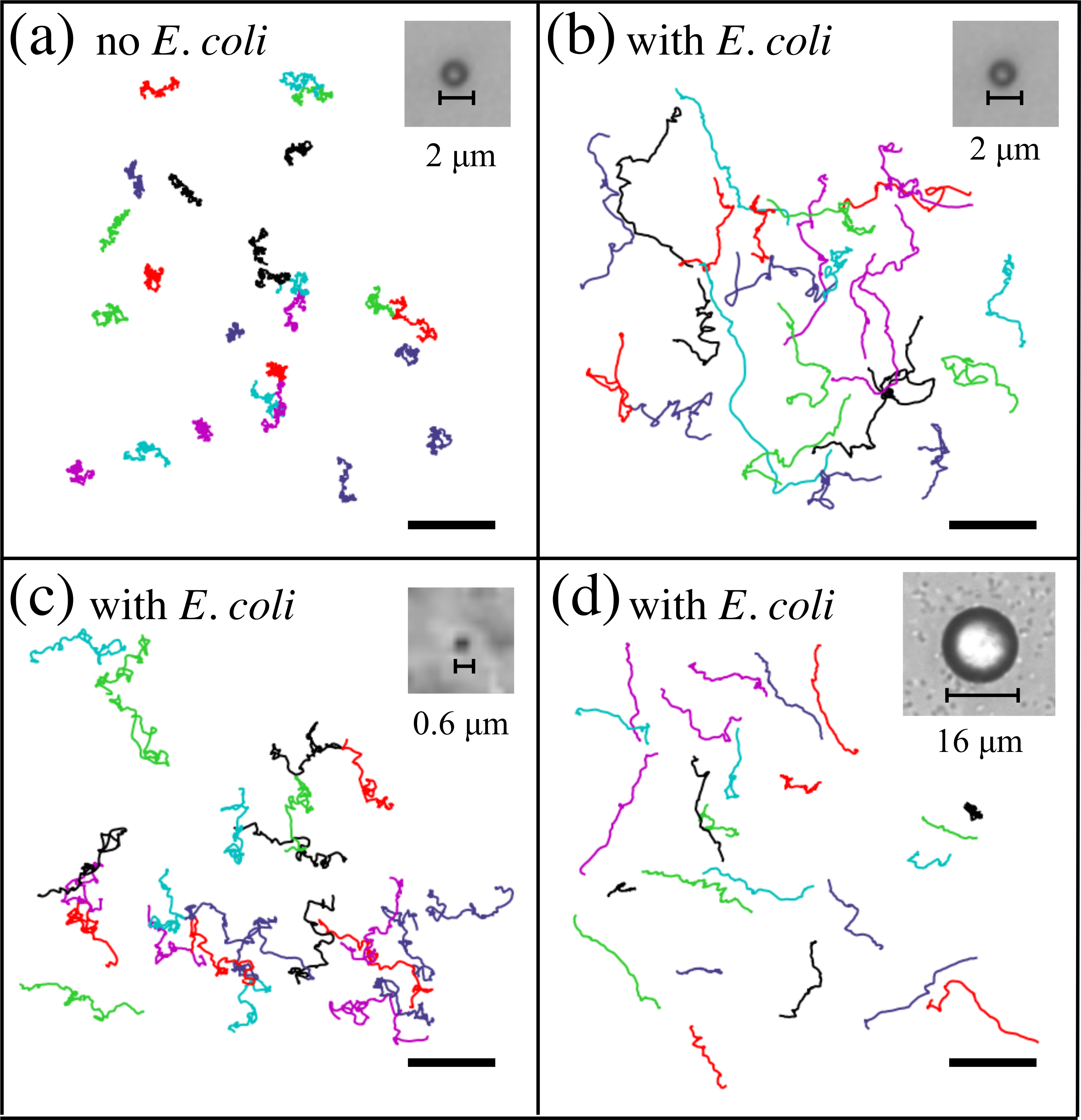}
  \caption{(Color online) Trajectories of 2 $\mu$m particles (a) without bacteria and (b) with bacteria ($c=3 \times 10^{9}$ cells/mL) for time interval 8 s. Trajectories of (c) 0.6 and (d) 16 $\mu$m particles ($c=3 \times 10^{9}$ cells/mL). Scale bar is 20 $\mu$m.}
  \label{fgr:example}
\end{figure}

The diffusion of molecules and particles in a fluid is a process that permeates many aspects of our lives including fog formation in rain or snow \cite{Klett1978}, cellular respiration \cite{Perry2007}, and chemical distillation processes \cite{Gorak2014}. At equilibrium, the diffusion of colloidal particles in a fluid is driven by thermal motion and damped by viscous resistance~\cite{Einstein1905}. In non-equilibrium systems, fluctuations are no longer only thermal and the link between these fluctuations and particle dynamics remain elusive\cite{durian2004}. Much effort has been devoted to understanding particle dynamics in non-equilibrium systems, such as glassy materials and sheared granular matter~\cite{Ramos2005}. A non-equilibrium system of emerging interest is active matter. Active matter includes active fluids, that is, fluids that contain self-propelling particles, such as motile microorganisms \cite{Sokolov2009,Zhang2010}, catalytic colloids \cite{Takagi2013,Palacci2013} and molecular motors~\cite{Leibler1997}. These particles inject energy, generate mechanical stresses, and create flows within the fluid medium even in the absence of external forcing~\cite{Ramaswamy2010,Marchetti2012}. Consequently, active fluids display fascinating phenomena not seen in  passive fluids, such as spontaneous flows~\cite{Zhang2010}, anomalous shear viscosities \cite{Sokolov2009,Mussler2013}, unusual polymer swelling \cite{Lowen2014,Patteson2015}, and enhanced fluid mixing \cite{Kim2004,Leptos2009,Childress2010,Kurtuldu2011}. Active fluids also play important roles in varied biological and ecological settings, which include the contributions of suspensions of microorganisms to biofilm infections \cite{Costerton1999,Josenhans2002}, biofouling of water-treatment systems \cite{Bhushan2012}, and biodegradation of environmental pollutants \cite{Valentine2010}.

The motion of passive particles in active fluids (e.g. suspension of swimming microorganisms) can be used to investigate the non-equilibrium properties of such fluids. For example, at short times particle displacement distributions can exhibit extended non-Gaussian tails as seen in experiments with algae \emph{C. reinhardtii} \cite{Leptos2009,Kurtuldu2011}. At long times, particles exhibit enhanced diffusivities $D_{\mathrm{eff}}$ greater than $D_0$, their thermal (Brownian) diffusivity\cite{Wu2000,Kim2004,Leptos2009,Childress2010,Mino2011,Kurtuldu2011,Jepson2013}. These traits are a signature of the non-equilibrium nature of active fluids; the deviation from equilibrium also manifests in violations of the fluctuation dissipation theorem \cite{Chen2007}.

In bacterial suspensions, the enhanced diffusivity $D_{\mathrm{eff}}$ depends on the concentration $c$ of bacteria. In their seminal work,  Wu and Libchaber \cite{Wu2000} experimentally found that $D_{\mathrm{eff}}$ increased linearly with $c$ in suspensions of \textit{E. coli}. Subsequent studies~\cite{Graham2008,Graham2009,Mino2011,Jepson2013,Yeomans2013,Kasyap2014} have observed that this scaling holds at low concentrations and in the absence of collective motion. In this regime, $D_{\mathrm{eff}}$ can be decomposed into additive components as $D_{\mathrm{eff}}=D_{0}+D_{\mathrm{A}}$~\cite{Graham2008,Graham2009,Mino2011,Jepson2013,Yeomans2013,Kasyap2014} where 
$D_{0}$ and $D_{\mathrm{A}}$ are the thermal and active diffusivities, respectively. It has been proposed that the active diffusivity $D_{\mathrm{A}}$ is a consequence of advection due to far-field interactions with bacteria \cite{Jepson2013} and may even be higher near walls \cite{Mino2011,Jepson2013}. 
  
While a majority of studies have focused on the role of bacterial concentration $c$ on particle diffusion, the role of particle diameter $d$ remains unclear. In the absence of bacteria, the diffusivity of a sphere follows the Stokes-Einstein relation, $D_{0} = k_{B}T /f_{0}$, where $k_{B}$ is the Boltzmann constant, $T$ is the temperature, and $f_{0} = 3 \pi \mu d$ is the Stokes friction factor\cite{Einstein1905} in a fluid of viscosity $\mu$. In a bacterial suspension, this relation is no longer expected to be valid. Surprisingly, for large particles (4.5 and 10 $\mu$m), Wu and Libchaber \cite{Wu2000} suggested that $D_{\mathrm{eff}}$ scales as $1/d$, as in passive fluids. Recent theory and simulation by Kasyap \emph{et al.} \cite{Kasyap2014} however do not support the $1/d$ scaling and instead predict a non-trivial dependence of $D_{\mathrm{eff}}$ on particle size, including a peak in $D_{\mathrm{eff}}$. This non-monotonic dependence of $D_{\mathrm{eff}}$ on particle size implies that measures of effective diffusivities \cite{Wu2000}, effective temperatures~\cite{Loi2008,Marchetti2012}, and momentum flux \cite{Mino2011,Jepson2013} intimately depend on the probe size and thus are not universal measures of activity. This has important implications for the common use of colloidal probes in gauging and characterizing the activity of living materials, {such as suspensions of bacteria \cite{Mino2011, Jepson2013}, biofilms \cite{Liana2015,Rogers2008}, and the cytoskeletal network inside cells \cite{Fodor2015}}, as well as in understanding transport in these biophysical setting. Despite the ubiquity of passive particles in active environments, the effects of size on particle dynamics in active fluids has yet to be systematically investigated in experiments. 

\begin{figure}[h]
\centering
  \includegraphics[width=\columnwidth]{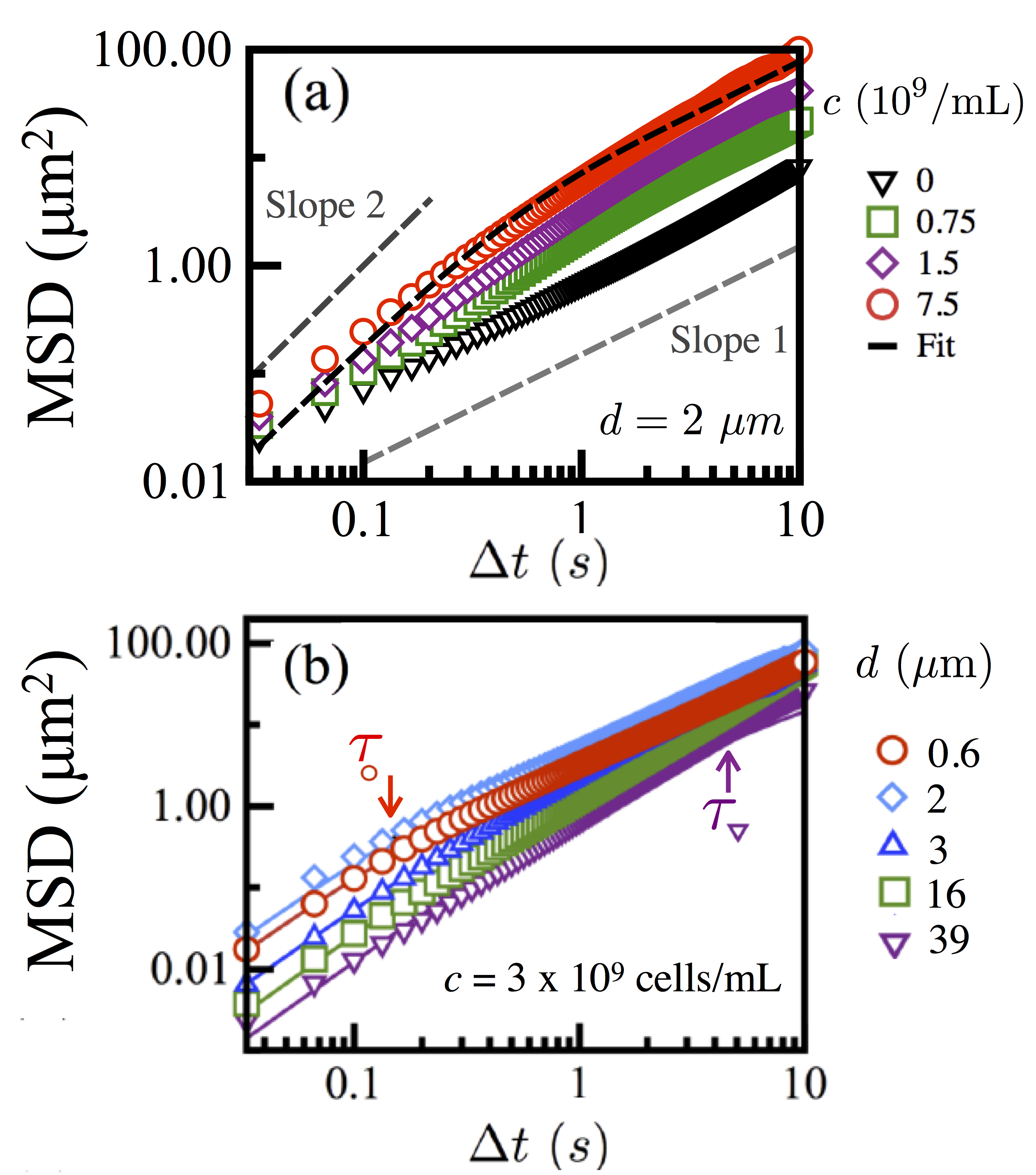}
  \caption{(Color online)(a) Mean-square displacements (MSD) of 2 $\mu$m particles versus time $\Delta t$ for varying bacterial concentration $c$. Dashed line is a fit to Langevin dynamics (eqn (1)). (b) MSD for varying particle diameter $d$ versus time for bacterial concentration $c= 3.0 \times 10^9$ cells/mL. The MSD peaks at $d=$ 2 $\mu$m. The cross-over time $\tau$ (arrows) increases with $d$. Solid lines are Langevin dynamics fits.}
\label{fgr:example}
\end{figure}

In this manuscript, we experimentally investigate the effects of particle size $d$ on the dynamics of passive particles in suspensions of {\it Escherichia coli}. {\it Escherichia coli} \cite{Berg2008} are model organisms for bacterial studies and are rod-shaped cells with 3 to 4 flagella that bundle together as the cell swims forward at speed $U$ approximately 10 $\mu$/s. We change the particle size $d$ from 0.6 $\mu$m to 39 $\mu$m, above and below the effective total length ($L \approx 7.6$ $\mu$m) of the {\it E. coli} body and flagellar bundle. We find that $D_{\mathrm{eff}}$ is non-monotonic in $d$, with a peak at $2<d<10~\mu$m; this non-monotonicity is unlike the previously found $1/d$ scaling \cite{Wu2000} and suggests that larger particles can diffuse faster than smaller particles in active fluids. Furthermore, the existence and position of the peak can be tuned by varying the bacterial concentration $c$. The active diffusion $D_{\mathrm{A}}=D_{\mathrm{eff}} - D_{0}$ is also a non-monotonic function of $d$ and can be collapsed into a master curve when rescaled by the quantity $cUL^4$ and plotted as a function of the P\'{e}clet number Pe $=UL/D_{0}$ (\emph{cf.} Fig. 5(b)). This result suggests that the active contribution to particle diffusion can be encapsulated by an universal dimensionless dispersivity $\bar{D}_{\mathrm{A}}$ that is set by the ratio of times for the particle to thermally diffuse a distance $L$ and a bacterium to swim a distance $L$.

\section{Experimental Methods}

\begin{figure*}[]
\centering
  \includegraphics[width=1.8\columnwidth]{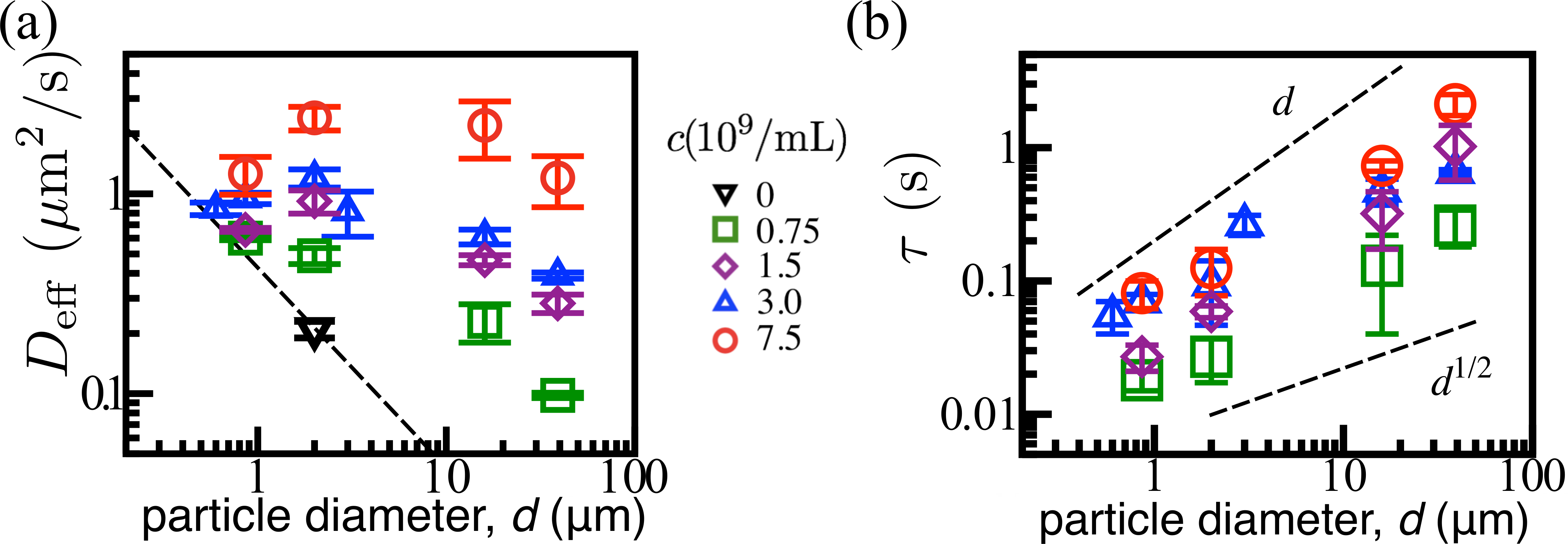}
  \caption{(Color online) (a) Effective particle diffusivities $D_{\mathrm{eff}}$ versus particle diameter $d$ at varying $c$. The dashed line is particle thermal diffusivity $D_0$. (b) The crossover-time $\tau$ increases with $d$, scaling as approximately $d^n$, where $1/2 \lesssim n \lesssim 1$. This is not the scaling in passive fluids \cite{Pathria1996} where $\tau \propto d^2$.}
  \label{fgr:example}
\end{figure*}

Active fluids are prepared by suspending spherical {polystyrene} particles and swimming \emph{E. coli} (wild-type K12 MG1655) in a buffer solution (67 mM NaCl in water). The \emph{E. coli}  are prepared by growing the cells  to saturation ($10^9$ cells/mL) in culture media (LB broth, Sigma-Aldrich). The saturated culture is gently cleaned by centrifugation and resuspended in the buffer. The polystyrene particles (density $\rho = 1.05$ g/cm$^3$) are cleaned by centrifugation and then suspended in the buffer-bacterial suspension, with a small amount of surfactant (Tween 20, 0.03\% by volume). The particle volume fractions $\phi$ are below 0.1\% and thus considered dilute. The \emph{E. coli} concentration $c$ ranges from 0.75 to 7.5 $\times 10^9$ cells/mL. These concentrations are also considered dilute, corresponding to volume fractions $\phi = c v_{\mathrm{b}} <$1\%, where $v_{\mathrm{b}}$ is the volume~\cite{Jepson2013} of an \emph{E. coli} cell body (1.4 $\mu$m$^3$). 

A 2 $\mu$l drop of the bacteria-particle suspension is stretched into a fluid film using an adjustable wire frame~\cite{Sokolov2009,Kurtuldu2011, Qin2015} to a measured thickness of approximately 100 $\mu$m. The film interfaces are stress-free, which minimizes velocity gradients transverse to the film. {We do not observe any large scale collective behavior in these films}; the \textit{E. coli} concentrations we use are below the values for which collective motion is typically observed \cite{Kasyap2014} ($\approx 10^{10}$ cells/ml ). Particles of different diameters (0.6~$\mu$m $<d<39~\mu$m) are imaged in a quasi two-dimensional slice (10 $\mu$m depth of focus). {We consider the effects of particle sedimentation, interface deformation, and confinement on particle diffusion and find that they do not significantly affect our measurements of effective diffusivity in the presence of bacteria (for more details, please see SI-$\S$I).} Images are taken at 30 frames per second {using a 10X objective (NA 0.45) and a CCD camera (Sony XCDSX90)}, which is high enough to observe correlated motion of the particles in the presence of bacteria (Fig. 2) but small enough to resolve spatial displacements. Particles less than $2$ $\mu$m in diameter are imaged with fluorescence microscopy (red, 589 nm) to clearly visualize particles distinct from \emph{E. coli} (2 $\mu$m long). We obtain the particle positions in two dimensions over time using particle tracking methods~\cite{Crocker1996,Qin2015}. All experiments are performed at $T_{0}=22^{\circ}$C.

\section{Results and Discussion}

\subsection{Mean Square Displacements}

Representative trajectories of passive particles in the absence and presence of \emph{E. coli} are shown in Fig. 1 for a time interval of 8 s. By comparing Fig. 1(a) (no \emph{E. coli}) to Fig. 1(b) ($c=3 \times 10^{9}$ cells/mL), we readily observe that the presence of bacteria enhances the magnitude of particle displacements compared to thermal equilibrium. Next, we compare sample trajectories of passive particles of different sizes $d$, below and above the \emph{E. coli} total length $L \approx 7.6~\mu$m. Figure 1(c) and 1(d) show the magnitude of particle displacement for $d=0.6~\mu$m and $d=16~\mu$m, respectively. Surprisingly, we find that the particle mean square displacements in the \emph{E. coli} suspension are relatively similar for the two particle sizes even though the thermal diffusivity $D_{0}$ of the 0.6~$\mu$m particle is 35 times larger than that of the 16~$\mu$m particle. The 16~$\mu$m particles also appear to be correlated for longer times than the 0.6~$\mu$m particles. These observations point to a non-trivial dependence of particle diffusivity on $d$  and are illustrated in sample videos included in the Supplementary Materials.

To quantify the above observations, we measure the mean-squared displacement (MSD) of the passive particles for varying  \emph{E. coli} concentration $c$ (Fig. 2(a)) and particle size $d$ (Fig. 2(b)). Here, we define the mean-squared particle displacement as ${\mathrm{MSD}}(\Delta t)=\langle |{\bf r}(t_{\mathrm{R}}+\Delta t)-{\bf r}(t_{\mathrm{R}})|^2 \rangle $, where the brackets denote an ensemble average over  particles and reference times $t_{\mathrm{R}}$. For a particle executing a random walk in two dimensions, the MSD exhibits a characteristic cross-over time $\tau$, corresponding to the transition from an initially ballistic regime for $\Delta t \ll \tau$ 
to a diffusive regime 
with ${\mathrm{MSD}} \sim 4 D_{\mathrm{eff}} \Delta t$ 
for $\Delta t \gg \tau$.

Figure 2(a) shows the MSD data for 2 $\mu$m particles at varying bacterial concentrations $c$. In the absence of bacteria ($c= 0$ cells/mL), the fluid is at equilibrium and $D_{\mathrm{eff}}=D_{0}$. For this case, we are unable to capture the crossover from ballistic to diffusive dynamics due to the lack of resolution: for colloidal particles in water, for example, cross-over times are on the order of nanoseconds and challenging to measure \cite{Florin2011}. Experimentally, the dynamics of passive particles at equilibrium are thus generally diffusive at all observable time scales. We fit the MSD data for the $d = 2~\mu$m case (with no bacteria) to the expression ${\mathrm{MSD}} = 4 D_{\mathrm{0}} \Delta t$, and find that $D_{\mathrm{0}}\approx0.2$ $\mu$m$^2$/s. This matches the theoretically predicted value from the Stokes-Einstein relation; the agreement ($\triangledown$) can be visually inspected in Fig. 3(a).

In the presence of \emph{E. coli}, the MSD curves exhibit a ballistic to diffusive transition, and we find that the cross-over time $\tau$ increases with $c$. For $\Delta t \gg \tau$, the MSD $\sim \Delta t$ with a long-time slope that increases with bacteria concentration $c$. Additionally, the distribution of particle displacements follows a Gaussian distribution (see SI-$\S$II A for details and measures of the non-Gaussian parameter). These features, MSD $\sim \Delta t$ and Gaussian displacements, indicate that the long-time dynamics of the particles in the presence of \emph{E. coli} is diffusive and can be captured by a physically meaningful effective diffusion coefficient $D_{\mathrm{eff}}$. 

We next turn our attention to the effects of particle size. For varying particle diameter $d$ at a fixed bacterial concentration ($c=3 \times 10^{9}$ cells/ml), the MSD curves also exhibit a ballistic to diffusive transition, as shown in Fig. 2(b). Surprisingly, we find a non-monotonic behavior with $d$. For example, the MSD curve for the 2~$\mu$m case sits higher than the 39~$\mu$m case and the 0.6~$\mu$m case. This trend is not consistent with classical diffusion in which MSD curves are expected to decrease monotonically with $d$ ($D_{0}\propto1/d$). We also observe that the cross-over time $\tau$ increases monotonically with $d$. As we will discuss later in the manuscript, the cross-over time scaling with $d$ also deviates from classical diffusion.

\begin{figure*}
\centering
  \includegraphics[width=2\columnwidth]{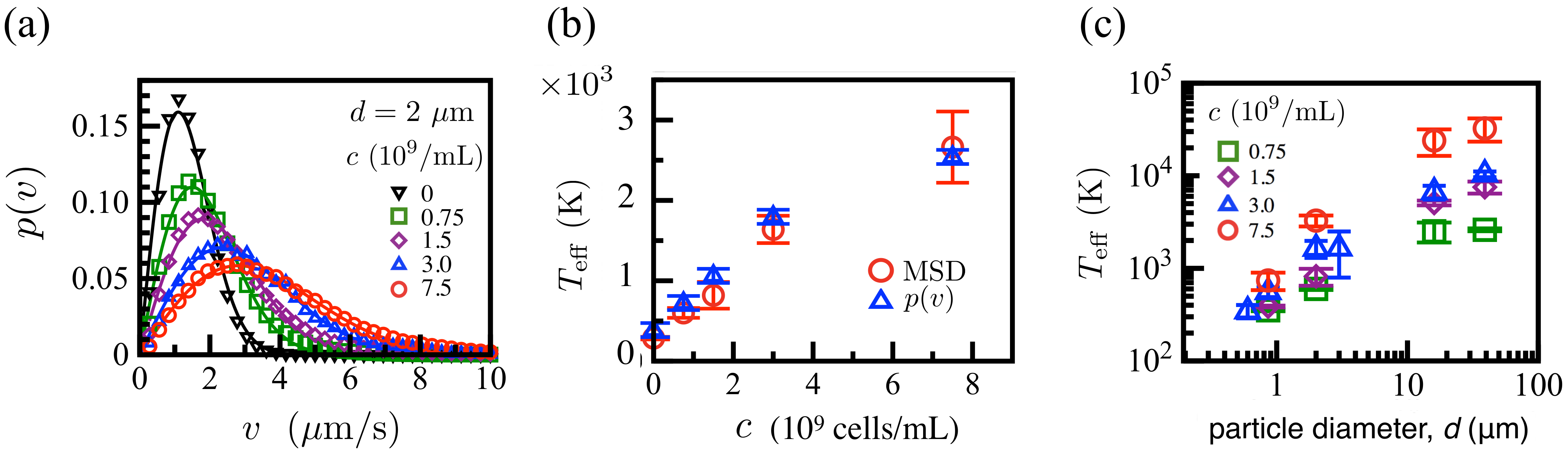}
  \caption{(Color online) (a) Distribution of 2 $\mu$m particle speeds $p(v)$ follow a Maxwell-Boltzmann distribution (solid curves) with clear peaks that shift right as $c$ increases. (b) The effective temperature $T_{\mathrm{eff}}$ extracted by fitting $p(v)$ data to eqn (2) (\textcolor{blue}{ $\bigtriangleup$}) matches those obtained from an extended Stokes-Einstein relation (\textcolor{red}{$\bigcirc$}). (c) The effective temperature $T_{\mathrm{eff}}$ increases with $d$ for varying $c$.}
  \label{fgr:example}
\end{figure*}

\subsection{Diffusivity and Cross-over Times}

{We now estimate the effective diffusivities $D_{\mathrm{eff}}$ and cross-over times $\tau$ of the passive particles in our bacterial suspensions. To obtain $D_{\mathrm{eff}} $ and $\tau$, we fit the MSD data shown in Fig. 2 to the MSD expression attained from the generalized Langevin equation\cite{Pathria1996}, that is
\begin{equation}
{\mathrm{MSD}}(\Delta t)= 4 D_{\mathrm{eff}} \Delta t \left( 1-{\tau \over \Delta t} \left(1-e^{-{\Delta t \over \tau}}\right) \right).
\end{equation} 
Equation~1 has been used previously to interpret the diffusion of bacteria \cite{Patteson2015} as well as the diffusion of particles in films with bacteria\cite{Liana2015}. In the limit of zero bacterial concentration, $D_{\mathrm{A}} =0$ and eqn (1) reduces to the formal solution to the Langevin equation for passive fluids, ${\mathrm{MSD}}(\Delta t)= 4 D_{0} \Delta t \left( 1-{\tau_{0} \over \Delta t} \left(1-e^{-{\Delta t / \tau_{0}}}\right) \right)$ with $\tau_{0} = \tau(c=0)$. For more details on the choice of our model, see SI-$\S$III.}

Figure 3 shows the long-time particle diffusivity $D_{\mathrm{eff}}$ (Fig. 3(a)) and the cross-over time $\tau$ (Fig. 3(b)) as a function of $d$ for bacterial concentrations $c = 0.75$, $1.5$, $3.0$ and $7.5 \times 10^9$ cells/mL. We find that for all values of $d$ and $c$ considered here, $D_{\mathrm{eff}}$ is larger than the Stokes-Einstein values $D_0$ at equilibrium (dashed line). For the smallest particle diameter case ($d=0.6$ $\mu$m), $D_{\mathrm{eff}}$ nearly matches $D_0$. This suggests that activity-enhanced transport of small ($d \lesssim 0.6$ $\mu$m) particles or molecules such as oxygen, a nutrient for \textit{E. coli}, may be entirely negligible~\cite{Kasyap2014}. For more information, including figures illustrating the dependence of $D_{\mathrm{eff}}$ on $c$ and comparisons between our measured effective diffusivities and previous experimental work, see sections SI-$\S$IIB and SI-$\S$IIC in the supplemental materials.

Figure 3(a) also reveals a striking feature, a peak $D_{\mathrm{eff}}$ in $d$.  Our data demonstrates that, remarkably, larger particles can diffuse faster than smaller particles in suspensions of bacteria. For example, at $c=7.5 \times 10^{9}$ cells/mL (\textcolor{red}{$\bigcirc$}) the $2$ $\mu$m particle has an effective diffusivity of approximately 2.0 $\mu$m$^{2}/s$, which is nearly twice as high as the effective diffusivity of the 0.86 $\mu$m particle, $D_{\mathrm{eff}}=$1.3 $\mu$m$^{2}/$s. We also observe that the peak vanishes as $c$ decreases. In fact, for the lowest bacterial concentration ($c=0.75 \times 10^9$ cells/mL), there is no peak: $D_{\mathrm{eff}}$ decreases monotonically with $d$. Clearly, $D_{\mathrm{eff}}$ does not scale as $1/d$.

Figure 3(b) shows the cross-over times $\tau$ characterizing the transition from ballistic to diffusive regimes as a function of particle size $d$ for varying $c$. We find that the values of $\tau$ increase with $d$ and $c$. We note that the variation of $\tau$ with $c$ (SI-Fig. 3(b)) does not follow a linear form. Instead, the data suggests possible saturation of $\tau$ for suspensions of higher -- but still dilute -- concentrations. The cross-over time $\tau$ scales with particle diameter approximately as $\tau \sim d^n$, where ${1 \over 2} \lesssim n \lesssim 1$. This cross-over time does not correspond to the inertial relaxation of the particle. Therefore, this scaling does not follow the trend seen for passive particles at thermal equilibrium \cite{Pathria1996}, where $\tau$, being the Stokes relaxation time (order 1 ns), scales as $m/f_0\propto d^2$ with particle mass $m \propto d^3$. Our data (Fig. 3(b)) highlights that in active fluids the super-diffusive motion of the passive particles cannot be ignored -- even for time scales as large as a second -- and that the time scales over which diffusive motion is valid ($\Delta t > \tau$) depends on the size $d$ of the particle. Further implications of a particle size-dependent cross-over time will be discussed below.

\subsection{Effective Temperature}

Our data so far suggests that particle dynamics in bacterial suspensions, while having an anomalous size-dependence (Figs. 1,2,3), maintain the characteristic super-diffusive to diffusive dynamics for passive fluids \cite{Pathria1996}. The long-time diffusive behavior (Fig. 2) and enhancement in $D_{\mathrm{eff}}$ (Fig. 3(a)), which is rooted in particle-bacteria encounters, suggest that the particles behave as if they are suspended in a fluid with an effectively higher temperature. 

In order to further explore the concept of effective temperature in bacterial suspensions, we measure the distribution of particle speeds $p(v)$ as a function of bacterial concentration $c$. The particle speed distributions determine the mean kinetic energy of the particles. If the distribution follows a Maxwell Boltzmann form -- as is always the case in fluids at equilibrium -- the mean kinetic energy is related to the thermodynamic temperature via the equipartition theorem. Such a relationship may not always exist for out-of-equilibrium fluids.

Figure 4(a) shows $p(v)$ for $d=2$ $\mu$m case for a range of $c$. We define the particle speeds $v$ over a time interval of $0.5$ s. This time interval is greater than the crossover times for the 2 $\mu$m particles (\emph{cf}. SI-Fig. 3) to ensure that a particle samples multiple interactions with bacteria and exhibits diffusive behavior. In the absence of bacteria, the system is in thermal equilibrium and $p(v)$ follows the two dimensional Maxwell-Boltzmann distribution,
\begin{equation}
p(v) = v m ( k_{B}T_{\mathrm{eff}})^{-1} {\mathrm{e}}^ {{-mv^2 \over 2k_{B}T_{\mathrm{eff}}}},
\end{equation}  
with peak speeds $v_{\mathrm{max}}=\sqrt{2k_{B}T_{\mathrm{eff}}/m}$, where $m$ is the mass of the polystyrene particle. Fitting the $p(v)$ data in the absence of bacteria ($\nabla$ in Fig. 4(a)) to eqn (2) yields $T_\mathrm{{eff}} \approx T_0$, as expected.

In the presence of bacteria, the particle speed distributions also follow a Maxwell-Boltzmann form, eqn (2), with peaks that shift toward higher values of $v$ as the \emph{E. coli} concentration $c$ increases. {Because our data follows the Maxwell-Boltzmann form (Fig. 4(a)) for all $c$, this indicates that there is no correlated motion at long times as is the case in swarming bacterial suspensions, for which the particle speed distribtuions exhibits an exponential decay \cite{Sokolov2010}.} We also note that the power spectra of particle speeds (SI-$\S$II D) are reasonably flat, consistent with white-noise forcing and an absence of correlated motion. Figure 4(a) thus indicates that an effective temperature $T_{\mathrm{eff}}$ can be defined from $p(v)$ and is increasing with the bacterial concentration. 

To quantify this `enhanced' temperature and the deviation from equilibrium behavior, we fit the $p(v)$ data (Fig. 4(a)) to eqn (2). These fits allow us to obtain $T_{\mathrm{eff}}$ as a function of $c$, as shown in Fig. 4(b) (\textcolor{blue}{ $\bigtriangleup$}). Also shown in Fig. 4(b) are the values extracted of $T_{\mathrm{eff}}$ (\textcolor{red}{ $\bigcirc$}) from an extended Stokes-Einstein relation, $T_{\mathrm{eff}}= 3 \pi \mu d D_{\mathrm{eff}}/k_B$, where $D_{\mathrm{eff}}$ are the values from the MSD fits shown in Fig. 3(a) and $\mu$ is the viscosity of the solvent ($\mu \approx 1$ mPa s). We find that both estimates of $T_{\mathrm{eff}}$ are higher than room temperature $T_0$ and increase linearly with $c$. 

{For a fluid at equilibrium, the temperatures from these two methods, MSD and $p(v)$, are expected to be the same but not for systems that are out of equilibrium \cite{Pathria1996,durian2004,Shokef2011,Gov2015}, such as the bacterial suspensions investigated here.} The good agreement in Fig. 4(b) suggests that $T_{\mathrm{eff}}$ may be a useful signature of bacterial activity with some analogies to equilibrium systems. {Note that in defining the effective temperature via the generalized Stokes-Einstein relation, we have assumed the viscosity to be a constant and independent of bacterial concentration, which may not be true when the bacterial concentrations are sufficiently high \cite{Clement2015}. Figure 4(b) suggests that an unchanging viscosity is a valid assumption for our system.}

\begin{figure}[h]
\centering
  \includegraphics[width=.9\columnwidth]{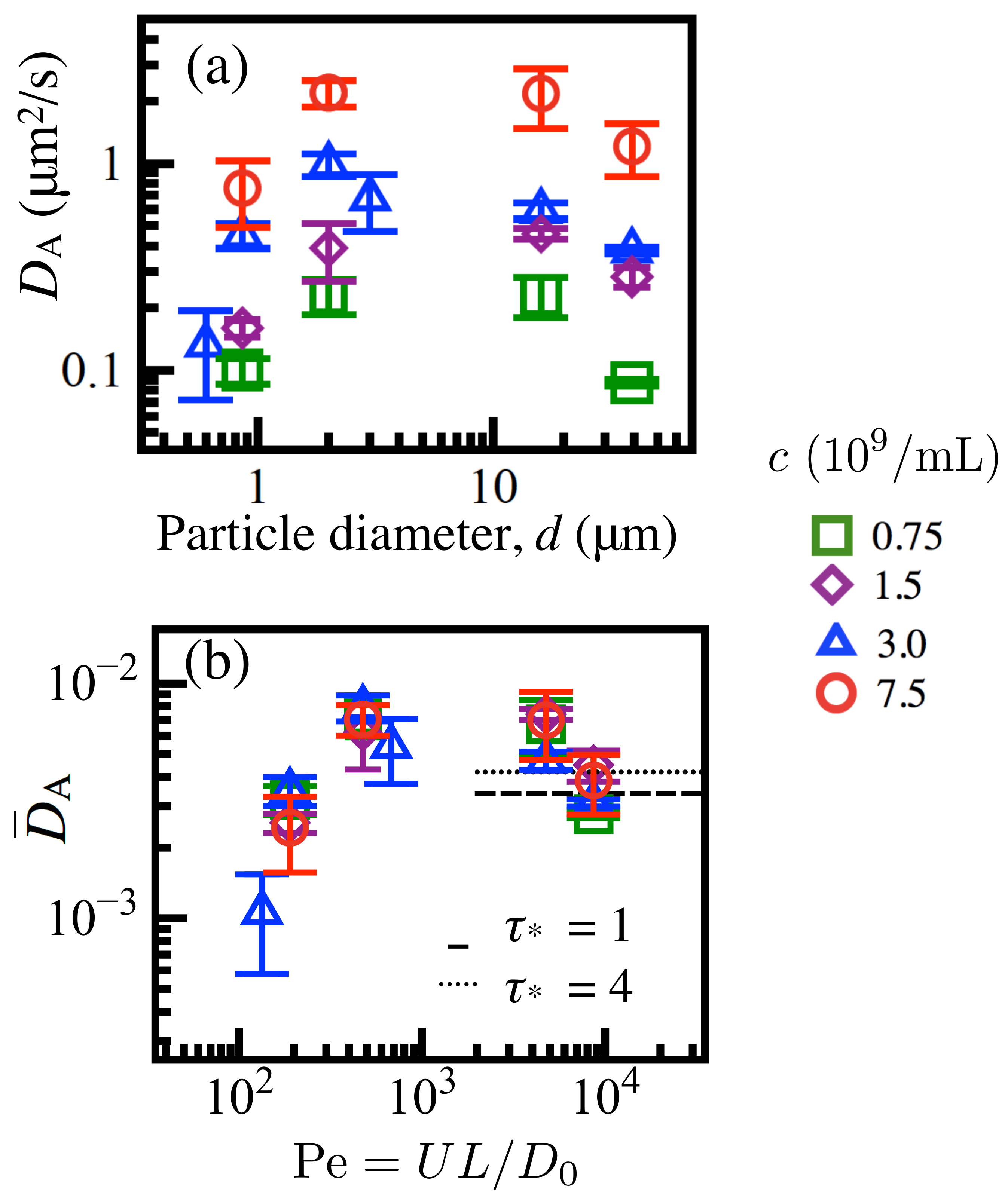}
  \caption{(Color online) (a) Active diffusivities $D_A = D_{\mathrm{eff}} - D_{0}$ are non-monotonic with particle size for varying concentrations of bacteria. (b) Scaled hydrodynamic diffusivity $\overline{D}_{\mathrm{A}} = D_A/ cUL^4$ collapses with P\'{e}clet number Pe = $UL/D_{\mathrm{0}}$. The maximum $\overline{D}_{\mathrm{A}}$ occurs at Pe$_{\mathrm{A}}$ between 450 and 4500. For $100 <$  Pe $<$ Pe$_{\mathrm{A}}$, $\overline{D}_{\mathrm{A}}$ scales as Pe$^{\alpha}$, where $\alpha \approx$ 2.}
  \label{fgr:example}
\end{figure}

Next, we investigate the role of particle diameter in the suspension effective temperature. Figure 4(c) show the values of $T_{\mathrm{eff}}$ estimated from an extended Stokes-Einstein relation as a function of $d$ for different values of $c$. Surprisingly, we find that $T_{\mathrm{eff}}$ increases with particle size $d$, which is clearly different from thermally equilibrated systems where temperature does not depend on the probe size. We note that for the largest particle diameter, $d=39~\mu$m, estimated $T_{\mathrm{eff}}$ values are approximately 100 times greater than room temperature $T_{0}=295$~K, consistent with previous reports~\cite{Wu2000}. {The dependence of $T_{\mathrm{eff}}$ on particle size $d$ (Fig. 5(a)) may be understood through the extended Stokes-Einstein relation, which yields $T_{\mathrm{eff}} = D_{\mathrm{eff}}f_0/k_B = T_{0} + \left(3 \pi \mu D_{\mathrm{A}}/k_B \right)d$, where $T_{0} = (3\pi\mu D_{0}/k_B)d$. If $D_{\mathrm{A}}$ were independent of $d$, then $T_{\mathrm{eff}}$ would be linear in $d$. However, a linear fit does not adequately capture the trend. This hints at a particle size dependent active diffusivity $D_{\mathrm{A}}$, which has been predicted in recent theory \cite{Kasyap2014}.} This variation of $T_{\mathrm{eff}}$ with $d$ highlights the interplay between particle size and the properties of the self-propelling particles (\emph{E. coli}) as well as challenges in gauging activity using passive particles.

\subsection{Active Diffusivity of Passive Particles in Bacterial Suspensions}

To explore the aforementioned dependence of $D_{\mathrm{A}}$ on particle size, we plot $D_{\mathrm{A}}= D_{\mathrm{eff}} - D_0$ with $d$. {Here, $D_0$ is the particle thermal diffusivity (in the absence of bacteria) and is obtained from the Stokes-Einstein relation $D_0 = {{k_BT_{0}} \over {3 \pi \mu d}}$.} Indeed, as shown in Fig. 5(a), $D_{\mathrm{A}}$ exhibits a non-monotonic dependence on $d$ for all $c$.

{To understand the observed dependence of the active diffusivity on the particle size, we consider the relevant time scales in our particle/bacteria suspensions, namely: (i) the time for the particle to thermally diffuse a distance $L$ equal to the total bacterial length $L^{2}/D_{0}$, (ii) the time for a bacterium to swim at a speed $U$ for a distance $L$ given by 
$L/U$, and (iii) the mean run time which is the inverse of the tumbling frequency $\omega^{-1}_{T}$.  Dimensionless analysis then suggests there are two independent time parameters: (i) the ratio of the first two above, which is the P\'{e}clet number, Pe $\equiv UL/D_{0}$, and (ii) $\tau^{*}=U/\omega_T L$, which is the ratio of the run length $U/\omega_{T}$ to the bacterial length.}

{It is these two parameters, Pe and $\tau^{*}$, that govern the particle dynamics in our experiments. When the P\'{e}clet number is much less than one, then the thermal particle diffusion dominates and transport by the bacteria is ineffective. When the P\'{e}clet number is much larger than one, then thermal diffusion is negligible and the transport is due to the convection from bacteria.The swimming speed $U$, tumble frequencies $\omega^{-1}_{T}$, and combined length of the cell body and flagellar bundle $L$ are estimated from prior experiments \cite{Patteson2015, Min2009} with \emph{E. coli} as approximately 10 $\mu$m/s, 1 s$^{-1}$, and $L \approx 7.6$ $\mu$m, respectively. Thus, in our experiments, the P\'{e}clet number varies from approximately 130 to 8600, via the particle bare diffusivity $D_{0}$ (through the particle diameter). We note that one stain of bacteria is used -- thus, the run length and bacteria size do not change in our experiments, and consequently, $\tau^{*} \approx 1.8$ is a constant.}

In order to gain insight into the non-trivial dependence of $D_{A}$ on $d$, we scale out the concentration dependence by introducing the dimensionless active diffusivity $\overline{D}_{\mathrm{A}}=D_A/ cUL^4$ and plot it  against the P\'{e}clet number, Pe. Figure 5(b) shows that all the active diffusion $D_{A}$ versus $d$ data shown in Figure 5(a) collapses into a single master curve, thereby indicating that $\overline{D}_{\mathrm{A}}$ is independent of $c$, at least for dilute suspensions investigated here. 

{We find that for Pe $\lesssim 10^3$, the values of $\overline{D}_{\mathrm{A}}$ initially increases with increasing Pe (or particle size) and follows a scaling ${\overline{D}}_{A} \sim {\mathrm{Pe}}^{2}$.  The observed increase of $\overline{D}_{\mathrm{A}}$ with Pe may be due to the decreasing particle Brownian motion, which allows the particle's motion to be correlated with the bacterial velocity disturbances for longer times.}

{In the limit of Pe $\rightarrow \infty$, the particle's Brownian motion vanishes and the particle displacements are dominated by the convective transport via bacteria-particle interactions. Thus ${\overline{D}_{A}}$ is expected to be independent of Pe and depend only on the parameter $\tau^{*}$, defined here as the ratio of the run length $U/\omega_{T}$ to the bacterial length $L$. Given that our experiments correspond to a constant $\tau^{*} \approx 1.8$, our experimentally measured $\overline{D}_{A}$ agree well with recent theoretical predictions \cite{Kasyap2014} for very large P\'{e}clet (as Pe $\rightarrow \infty$) for $\tau^{*}=1$ and $\tau^{*}=4$, as shown in Fig. 5(b). An increase in the run time (or $\tau^{*}$) would increase the asymptotic value of the scaled active diffusivity ${\overline{D}_{A}}$ (see SI-$\S$IV).}

{An important feature of the data shown in Fig. 5(b) is the peak in ${\overline{D}_{A}}$ at Pe$_{\mathrm{A}} \approx 10^3$. The existence of such a peak has been predicted in a recent theory/simulation investigation \cite{Kasyap2014}, and our data agrees at least qualitatively with such predictions. We note that the predicted peak in ${\overline{D}_{A}}$ happens at a Pe $\approx O(10)$, which is smaller than our experimental values of (Pe~$\approx 10^3$). The appearance of the peak in ${\overline{D}_{A}}$ in our data may be due to the weak but non-zero effects of Brownian motion, which allows particles to sample the bacterial velocity field in such a way that the mean square particle displacements and correlation times are higher compared to both the very high Pe (negligible Brownian motion) as well as the very low Pe (Brownian dominated) \cite{Kasyap2014}. This feature (i.e. peak in  ${\overline{D}_{A}}$) is surprising because it suggests an optimum particle size for maximum particle diffusivity that is coupled to the activity of the bacteria.}

\section{Maximum Particle Effective Diffusivity $D_{\mathrm{eff}}$}
 
Our data (Fig. 5(b)) shows that the dimensionless active diffusion $\overline{D}_{\mathrm{A}}$ collapses unto a universal curve with a peak in Pe for all bacterial concentrations. Our data also shows that, with the exception of the lowest bacterial concentration $c$, the particle effective diffusivity $D_{\mathrm{eff}}$ exhibits a peak in $d$, which varies with $c$ (Fig. 3(a)). For instance, for $c = 1.5 \times 10^9$ cells/mL, the peak is at approximately $2$ $\mu$m, while for larger concentrations ($c=7.5 \times 10^9$ cells/mL), the peak (as obtained by fitting the data to a continuous function) shifts to higher values of $d$. This suggests that one can select the particle size which diffuses the most by tuning the bacteria concentration. 
 
In what follows (see also SI-$\S$V), we provide a prediction, based on our experimentally-measured universal curve of ${\overline{D}}_{A}$ with Pe (Fig. 5(b)) for the existence as well as the location of the peak of $D_{\mathrm{eff}}$ in $d$. As noted before, the particle effective diffusivity $D_{\mathrm{eff}}$ can be described as the linear sum of the particle thermal diffusivity $D_{\mathrm{0}}$, which is independent of $c$ and decreases with $d$, and the active diffusivity $D_{\mathrm{A}}$, which is linear in $c$ and non-monotonic in $d$, through the particle-size dependent ${\overline{D}}_{A}$. Therefore, we can recast the effective diffusivity as

\begin{equation}
D_{\mathrm{eff}} =D_{0} + (cL^{3}) \:(U L) {\overline{D}}_{A}.
\end{equation}

\noindent{}The criterion for the existence of a maximum $D_{\mathrm{eff}}$ is obtained by taking the derivative of eqn (3) with respect to the P\'{e}clet number and setting the derivative to zero. {In order to estimate $\overline{D}_{\mathrm{A}}$ (and its slope), we fit the data in Fig. 5(b) near the peak in the range $200 <$ Pe $< 4000$ with a second order polynomial equation. We find that a peak exists in $D_{\mathrm{eff}}$ if}

\begin{equation}
{cL^3\gtrsim 0.4.}
\end{equation}

\noindent{}{For the bacterial length used here $L = 7.6$ $\mu$m, this yields $c \approx 0.9 \times 10^9$ cells/mL, which is in quantitative agreement with the concentration range ($0.75 \times 10^9$ cells/mL $ <c<1.5 \times 10^9$ cells/mL) in which the peak in $D_{\mathrm{eff}}$ emerges in our data (Fig. 3(a)). As described in SI-$\S$V, we find that the location of the $D_{\mathrm{eff}}$ peak in $d$ here defined as $d_{\mathrm{eff}}^{\mathrm{max}}$ is given by}

\begin{equation}
{d_{\mathrm{eff}}^{\mathrm{max}} \approx { d_{\mathrm{A}}}\left [ 1 - \left({ 1 \over
{{5cL^{3} -2}}}\right) \right]}
\end{equation}

\noindent{}{where $d_{\mathrm{A}}$ corresponds to the P\'{e}clet number $\mathrm{Pe}_{\mathrm{A}}\approx 1000$ at which ${\overline{D}}_{A}$
 is maximum. For the \emph{E. coli} used here, $d_{\mathrm{A}}=kT\mathrm{Pe}_{\mathrm{A}}/ 3 \pi \mu UL \approx 6$ $\mu$m. Note that  $d_{\mathrm{eff}}^{\mathrm{max}}$ is always less than or equal to $d_{\mathrm{A}}$ and increases with $c$, consistent with our experimental observations (Fig. 3(a)). In general the criterion, eqn (4), and $d_{\mathrm{eff}}^{\mathrm{max}}$, eqn (5), will depend on $\tau^*$. For suspensions of bacteria, the universal curve of ${\overline{D}}_{A}$ informs when and where a peak in the particle diffusivity occurs.}

\section{Conclusions}

In summary, we find that the effective particle diffusivity $D_{\mathrm{eff}}$ and temperature $T_{\mathrm{eff}}$ in suspensions of \emph{E. coli} show strong deviations from classical Brownian motion in the way they depend on particle size $d$. For example, Fig. 3(a) shows that $D_{\mathrm{eff}}$ depends non-monotonically in $d$ and includes a regime in which larger particles can diffuse faster than smaller particles. {The existence as well as the position of a $D_{\mathrm{eff}}$ peak in $d$ can be tuned by varying the bacterial concentration $c$. We also find that the cross-over time $\tau$ increases with particle size and scales as approximately $d^n$, where $1/2 \lesssim n \lesssim 1$, as shown in Fig. 3(b).}

Measures of $T_{\mathrm{eff}}$ obtained from either an extended Stokes-Einstein relation or particle speed distributions seem to agree quite well (Fig. 4(b)). This is surprising since this kind of agreement is only expected for systems at equilibrium. The good agreement between the measurements suggests that $T_{\mathrm{eff}}$ may be a useful signature of bacterial activity. However, unlike thermally equilibrated systems, $T_{\mathrm{eff}}$ varies with size $d$ (Fig. 4(c)). This non-trivial dependence of both $D_{\mathrm{eff}}$ and $T_{\mathrm{eff}}$ on particle size $d$ implies that these common gauges of activity are not universal measures. Nevertheless, our data  suggest that one can define optimal colloidal probes of activity for suspensions of bacteria, which correspond to Pe $= UL/D_0 \approx 10^3$. At these Pe values, $\bar{D}_{\mathrm{A}}$ is maximized, which provides ample dynamical range (magnitude of the signal). Also at these Pe values, the cross-over time $\tau$ is still relatively small, which allows for adequate temporal resolution. Both of these features are important in using passive particles to characterize a spatially and temporally varying level of activity in materials. 

Our anomalous particle-size dependent results in active fluids has important implications for particle sorting in microfluidic devices, drug delivery to combat microbial infections, resuspension of impurities and the carbon cycle in geophysical settings populated by microorganisms. A natural next step would be to study the role of external fields such as gravity or shear in influencing particle transport in these active environments.

\section*{Acknowledgements}
\noindent{}We would like to thank J. Crocker, S. Farhadi, N. Keim, D. Wong, E. Hunter, and E. Steager for fruitful discussions and B. Qin for help in the spectral analysis of data. This work was supported by NSF-DMR-1104705 and NSF-CBET-1437482.

\footnotesize{
\bibliographystyle{rsc} 

}

\end{document}